\documentstyle[seceq,epsf,twoside]{ptptex}
\notypesetlogo  
\title{Property of Chiral Scalar and Axial-Vector Mesons in\\ 
       Heavy-Light Quark Systems}
\author{%
Muneyuki {\sc Ishida} and Shin {\sc Ishida}$^{*}$  }
\inst{%
Department of Physics, Tokyo Institute of Technology\\
Tokyo 152-8551, Japan\\
$^*$Atomic Energy Research Institute, 
College of Science and Technology\\
Nihon University, Tokyo 101-0062, Japan
}
\recdate{%
\today
}
\abst{%
Recently we have proposed 
a new level-classification scheme of hadrons with a manifestly covariant
framework.
In this scheme the requirement of chiral symmetry on the light quark leads to 
a prediction of
existence of new type of scalars $X_B,X_D$ and axial-vectors
$X_{B^*},X_{D^*}$ as the chiral partners of
ground state pseudoscalar $B,D$ and vector $B^*,D^*$ mesons,
respectively.
They belong to ``relativistic $S$-wave states,'' and are discriminated from 
the conventional $P$-wave mesons with $j_q=1/2$ 
appearing in the heavy quark effective theory.
In this talk we examine the properties of these chiral mesons:
The mass-splittings between the respective chiral partners are 
predicted to be equal,
and the decay widths of one pion emission of 
$X_B,X_D,X_{B^*}$ and $X_{D^*}$
are to take the same value due to both chiral and heavy quark symmetries.
Some experimental indications for 
existence of
$X_B$ and $X_{D^*}$ are also given, which
are consistent with the above prediction. 
}

\begin{document}
\maketitle

\setcounter{tocdepth}{4}

\section{Introduction}
Recently, we have proposed a covariant level classification 
scheme of hadrons\cite{rfCLC}, unifying the seemingly contradictory
two viewpoints,
non-relativistic one with
$LS$-coupling scheme 
and relativistic one with 
chiral symmetry.
In this scheme, it is expected that the hadron spectra are to show,
concerning light quark constituents, the approximate  
$\tilde U(12)_{SF}$ symmetry (including static $SU(6)_{SF}$ as a subgroup)
around the lower mass region,
and is predicted the existence of many relativistic states, called ``chiralons,"
which are out of the framework of conventional $LS$ coupling scheme in NRQM.  
Recently, the existence of a light scalar $\sigma$ meson with 
the property as partner of $\pi$ meson in the linear 
representation of chiral symmetry seems to be confirmed\cite{rfsigma,rf555}.
In our classification scheme,
this $\sigma$ is naturally classified into the relativistic 
$S$-wave $q\bar q$ state, which is to be discriminated from the
$^3P_0$-state appearing in NRQM.

In heavy-light quark $n\bar b(=u\bar b,d\bar b)$ systems, 
the existence of chiralons, new scalar and axial-vector mesons, 
denoted as $X_B=^t(X_{B^+},X_{B^0})$ and 
$X_{B^*}=^t(X_{B^{*+}},X_{B^{*0}})$, is expected to exist as 
the chiral partners of the pseudoscalar 
$B=^t(B^+,B^0)$ and the vector $B^*=^t(B^{*+},B^{*0})$ mesons, respectively.
Similarly, in the $n\bar c=^t(u\bar c, d\bar c)$ system, 
the scalar $X_{\bar D}=^t(X_{\bar D^0},X_{D^-})$ and 
the axial-vector $X_{\bar D^*}=^t(X_{\bar D^{*0}},X_{D^{*-}})$ mesons,
is expected to exist as 
the chiral partners of the pseudoscalar 
$\bar D=^t(\bar D^0,D^-)$ and the vector 
$\bar D^*=^t(\bar D^{*0}, D^{*-})$ mesons, respectively.
These chiral scalar and axial-vector mesons are 
classified to the
relativistic $S$-wave states, which are discriminated from the
$P$-wave mesons appearing in NRQM, or the scalars $B_0^*,D_0^*$ and 
axial-vectors $B_1^*,D_1^*$ with $j_q=1/2$ appearing in HQET. 

In this talk we investigate the properties of these chiral scalar 
and axial-vector mesons by taking into account 
chiral symmetry for the light quark component, as well as 
HQS for the heavy quark component.
A similar approach has already been done in 
Refs.~\cite{rfbardeen,rfebert}.
However, they assigned as the respective chiral partners of $S$-wave 
state mesons to the  $j_q=1/2$ 
$P$-wave mesons.
This assignment is crucially different from ours.

\section{Constraints\ on\ one\ pion\ emission\ process\ from\ HQS}
We consider general constraints from HQS on the processes of 
one-pion emission
$X_B\to B\pi$, $X_{B^*}\to B^*\pi$, $X_D\to D\pi$ and $X_{D^*}\to D^*\pi$,
which are expected to be the main decay modes of the relevant mesons. 

The HQ spin symmetry relates $B$, $X_B$, $D$ and $X_D$, respectively, 
to $B^*$, $X_{B^*}$, $D^*$ and $X_{D^*}$, and 
the HQ flavor symmetry relates $B$, $B^*$, $X_B$ and $X_{B^*}$, respectively,
to  $\bar D$, $\bar D^*$, $X_{\bar D}$ and $X_{\bar D^*}$ with the same velocity.
Thus, the $S$-matrix elements of the relevant four decay modes 
are related with one another, and are represented by one universal 
amplitude $\xi$ as.\footnote{ 
Here the normalization of states $|B({\bf 0})\rangle \equiv 
a_{B({\bf 0})}^\dagger |0\rangle$, etc. are used, where 
$[ a_{B({\bf p})}, a_{B({\bf p}')}^\dagger ]=\delta^{(3)}({\bf p}-{\bf p}') $.
}
\begin{eqnarray}
  &-i&
\langle \pi B^*({\bf 0},\epsilon^{(0)}) | U_I | X_{B^*}({\bf 0},\epsilon^{(0)})
 \rangle = 
  -i \langle \pi D^*({\bf 0},\epsilon^{(0)}) | U_I | X_{D^*}({\bf 0},
\epsilon^{(0)}) \rangle  \nonumber\\ 
 &=& \langle \pi B({\bf 0}) | U_I | X_B({\bf 0}) \rangle 
= \langle \pi D({\bf 0}) | U_I | X_D({\bf 0}) \rangle \nonumber\\
 &=&
 -\xi \frac{1}{(2\pi )^3} \sqrt{\frac{1}{(2\pi )^3 2 E_\pi } }
       i (2\pi )^4 \delta^{(4)}(P_{X_M}-P_M-p_\pi) ,
\label{eqhqs}
\end{eqnarray}
where $U_I$ is the translational operator of time 
from $-\infty$ to $+\infty$, $P_{X_M} (P_M)$ being the initial (final) heavy meson
momentum, $p_\pi$ being the emitted pion momentum,
and ${\bf 0}$ represents 
the three velocity ${\bf v}={\bf 0}$, 
the longitudinally polarized states of $B^*$, $D^*$ and $X_{B^*}$, 
$X_{D^*}$ appear, 
and a common factor of
$\frac{1}{(2\pi )^3} \sqrt{\frac{1}{(2\pi )^3 2 E_\pi } }$
has been introduced.

The decay widths of $X_M=X_B,X_{B^*},X_D,X_{D^*}$ are given by
\begin{eqnarray}
\Gamma_{X_M} &=&
 3 \frac{1}{2m_{X_M}} \frac{|{\bf p}|}{4\pi m_{X_M}}
 (2\xi \sqrt{m_{X_M}m_M})^2 
  \approx  3 \frac{\sqrt{(\Delta m_M)^2 -m_\pi^2}}{8\pi} (2\xi )^2 ,
\label{eqhqrel} 
\end{eqnarray}
where we use the approximation $m_{X_M}\approx m_M$, 
$|{\bf p}|=\sqrt{(\Delta m_M)^2-m_\pi^2}$ is the  
pion CM momentum, and 
the factor 3 comes from the isospin degree of freedom of the final $|\pi M \rangle$ state.
As expressed by Eq.~(\ref{eqhqrel}), the decay widths of the relevant 
processes are dependent only upon the corresponding mass difference
$\Delta m_M$.

\section{Chiral\ and\ Heavy\ Quark\ Symmetric\ Lagrangian}
Here we construct the chiral and heavy quark symmetric Lagrangian
of $B$ and $\bar D$ systems, by using 
the quark bi-spinor representation in the new 
classification scheme:
$
U^{(\pm )\beta}_\alpha (v)  = 
 \sum_\phi ({1}/{2\sqrt 2}) \Gamma_\phi \hat\phi (1+iv\cdot\gamma )_\alpha{}^\beta ,
$
where $\alpha (\beta)$ are light quark (heavy antiquark) spinor index, and 
the summation is taken for $\phi = B,X_B,B^*_\mu ,X_{B^*_\mu}$; 
$\bar D,X_{\bar D},\bar D^*_\mu ,X_{\bar D^*_\mu }$; 
$\Gamma_B=\Gamma_{\bar D}=i\gamma_5,$
$\Gamma_{X_B}=\Gamma_{X_{\bar D}}=\pm 1,$
$\Gamma_{B^*_\mu}=\Gamma_{\bar D^*_\mu}=i\gamma_\mu$, 
$\Gamma_{X_{B^*_\mu}}=\Gamma_{X_{\bar D^*_\mu}}=\pm \gamma_5 \gamma_\mu$.
The field 
$\hat\phi$ is related with the ordinarily normalized field $\phi$ as
$\phi (X)=({1}/{\sqrt{2m_Q}})e^{im_Qv\cdot X} \hat\phi (X)$, where $m_Q=m_b(m_c)$
for $B(\bar D)$ system. The chiral $U_A(1)$ transformation for the light quark is given by
$U^{(\pm )}\rightarrow e^{\pm i \alpha \gamma_5} U^{(\pm )}$. 
By using the conjugate bispinor, defined by 
$\bar U^{(\pm )}\equiv \gamma_4 U^{(\pm )\dagger} \gamma_4$, 
the free Lagrangian is given by 
\begin{eqnarray}
{\cal L}^{\rm free} &=& \langle \bar U^{(-)} 
\left( 
{iv\cdot \stackrel{\leftrightarrow }{\partial } }/{2}
-m_q 
\right)  
U^{(+)} \rangle
=\sum_\phi \hat\phi^\dagger 
\left( { iv\cdot \stackrel{\leftrightarrow}{\partial } }/{2}
-m_q\right)
\hat\phi , 
\end{eqnarray}
where $\langle\ \ \rangle$ means the trace on spinor indices, 
and the light quark mass $m_q$ is common for
all $B$ and $\bar D$ systems.
The total meson mass $m_{M}$ $(M$=$B,D)$ is given by $m_M$=$m_Q+m_q$,
thus  $m_B$=$m_{X_B}$=$m_{B^*}$=$m_{X_{B^*}}$=$m_b+m_q$ and  
 $m_D$=$m_{X_D}$=$m_{D^*}$=$m_{X_{D^*}}$\\
=$m_c+m_q$  
in symmetric limit.

The $\sigma$ and $\pi$ fields are transformed as
$(\sigma + i\gamma_5 \tau\cdot\pi )\rightarrow 
e^{i \alpha \gamma_5} (\sigma + i\gamma_5\tau\cdot\pi )e^{i \alpha \gamma_5} $,
thus the chiral symmetric Yukawa coupling is given by
\begin{eqnarray}
{\cal L}^{\rm Yukawa} &=& -\eta \langle \bar U^{(-)} 
(\sigma + i\gamma_5\tau\cdot\pi ) U^{(-)} \rangle .
\label{eqY}
\end{eqnarray}
Through the spontaneous breaking of chiral symmetry the $\sigma$ acquires 
vacuum expectation value 
$\langle\sigma\rangle \equiv\sigma_0(=f_\pi$ in SU(2)
linear $\sigma$ model), which induces the mass splittings $\Delta m$
between chiral partners. The $\Delta m$ are universal in 
$B$ and $\bar D$ systems:
\begin{eqnarray}
\Delta m &\equiv& m_{X_B}-m_B=m_{X_{B^*}}-m_{B^*}=  
 m_{X_D}-m_D=m_{X_{D^*}}-m_{D^*},
\end{eqnarray}
which is given by $\Delta m= 2\eta \sigma_0=2\eta f_\pi$ 
in Lagrangian~(\ref{eqY}).
Even in the case,
considering the contribution from all the other possible forms 
of effective chiral symmetric Lagrangian, 
the universality of mass-splittings 
are shown to be preserved.
Thus, following the argument given in the last sub-section, 
the decay widths of one pion emission
also become universal, 
\begin{eqnarray}
\Gamma (\Delta m) & \equiv & \Gamma_{X_B\rightarrow B\pi} = \Gamma_{X_D\rightarrow D\pi}
  =\Gamma_{X_{B^*}\rightarrow B^*\pi} = \Gamma_{X_{D^*}\rightarrow D^*\pi},
\end{eqnarray}
although the magnitude of $\Gamma (\Delta m)$
cannot be predicted in the present framework.
\footnote{
In Lagrangian~(\ref{eqY}), $\xi$ is given by  
$\xi =\eta =\Delta m/(2f_\pi)$. By taking $\Delta m\approx 300$MeV 
in Eq.~(\ref{eqhqrel}) as an example, 
 $\Gamma(\Delta m = 300$MeV)=331MeV.   
We can also consider, as one of the possible forms
of the effective Lagrangian, 
the ${\cal L}^{(d)}=
-k \langle \bar U^{(-)}[iv_\mu \partial_\mu (\sigma + i\tau\cdot\pi )] 
U^{(-)} \rangle$, where  $k$ is a coupling constant of 
${\cal O}(1/m_q)$. In this case 
$\xi = \eta + 2kv\cdot p_\pi \approx \eta - k\Delta m
=(1-2kf_\pi )\Delta m/(2f_\pi)$.
By taking a natural value of $k\approx 2/m_\rho$ as an example,
the $\Gamma(\Delta m = 300$MeV)=88 MeV. 
As is seen in these examples, the magnitude of $\Gamma (\Delta m)$ itself 
is largely dependent upon the value of $k$.}

\section{Experimental\ Evidence\ for\ $X_B$\ and\ $X_{D^*}$}
In order to examine phenomenologically 
whether these chiral mesons really exist or not,
we analyze the $B\pi$ mass spectra\cite{rfR1} 
in 5.4GeV $<m_{B\pi}<$ 5.9GeV, obtined
through $Z^0$-boson decay by L3\cite{L3} and ALEPH\cite{ALEPH} collaborations. 
In the relevant mass region of $B\pi$ channel, the $X_B$, and  
the $P$-wave mesons, $B_2^*(j_q=3/2)$ and $B_0^*(j_q=1/2)$,
are expected to be observed directly.
We use the following forms of squared amplitude $|{\cal M}|^2$ 
and background $|{\cal M}|^2_{\rm BG}$,
\begin{eqnarray}
|{\cal M}|^2 &=& |r_1e^{i\theta_1}\Delta_{X_B}(s)+r_2e^{i\theta_2}\Delta_{B_0^*}(s)|^2
  +|r_3e^{i\theta_3}\Delta_{{\rm other}\  B}(s)|^2  , \nonumber\\
|{\cal M}|^2_{\rm BG} &=& P_1(m_{B\pi}-P_2)^{P_3} 
e^{  P_4(m_{B\pi}-P_2)+P_5(m_{B\pi}-P_2)^2+P_6(m_{B\pi}-P_2)^3  } ,
\label{eqfit}
\end{eqnarray}
where $P_1$--$P_6$ are parameters and 
$\Delta_R(s)=-m_R\Gamma_R/(s-m_R^2+im_R\Gamma_R)$;
$\Delta_{X_B}(\Delta_{B_0^*})$ are the
Breit-Wigner amplitude 
for $X_B(B_0^*)$ mesons, and 
$\Delta_{{\rm other}\  B}$ represents contributions 
from all the other possible
resonances including $B_2^*$.\footnote{
In these experiments, the final photon was not detected, 
and so the following states decaying into $B^*\pi$ are also to be 
seen in $B\pi$ spectra indirectly through the successive $B^*\rightarrow B\gamma$ decay:
$B_2^*(\rightarrow B^*\pi \rightarrow \gamma B\pi )$,
$B_1(j_q=3/2)(\rightarrow B^*\pi \rightarrow \gamma B\pi )$,
$B_1^*(j_q=1/2)(\rightarrow B^*\pi \rightarrow \gamma B\pi )$ and  
$X_{B^*}(\rightarrow B^*\pi \rightarrow \gamma B\pi )$.
The observed $m_{B\pi}$-values of these resonances become smaller 
than their real values
by the missing photon energy $E_\gamma =m_{B^*}-m_B$.  
In Eq.~(\ref{eqfit}) $\Delta_{{\rm other}\   B}$ are meant as including, 
in addition to direct $B_2^*(\rightarrow B\pi )$, 
the above mentioned indirect $B_2^*(\rightarrow B^*\pi )$ and $B_1^*(\rightarrow B^*\pi)$,
which are $D$-wave decays. 
The indirect  $B_1^*(\rightarrow B^*\pi )$ and $X_{B^*}(\rightarrow B^*\pi)$,
which are $S$-wave decays interfering with each other, are considered 
to be included in $\Delta_{B_0^*}$ and $\Delta_{X_B}$, respectively.  
}  
Preliminary results of the fit are given in FIGURE {\bf 1}. 
Both data show a dip at the same energy $m_{B\pi}\approx 5.55$GeV,
which is reproduced by the interference between the $X_B$ with a narrow width and
the $B_0^*$ with a wide width. 
The mass and width of $X_B$ is given by
$m_{X_B} = 5540{\rm MeV}$ and  $\Gamma_{X_B}=21{\rm MeV}$,  
and the obtained values of $\chi^2$ is $\tilde\chi^2=22.92/20=1.15$.
We also tried the fit without $X_B$, and obtain almost the same value of 
$\tilde\chi^2=26.280/24=1.10$,  thus, only by this analysis,  
we cannot obtain the definite conclusion 
on existence of $X_B$.

Similar analysis is also done\cite{YAMADA} on the $D^*\pi$ mass spectra 
by CLEO and DELPHI
collaborations, and we obtained the preliminary result 
on existence of $X_{D^*}$ with
$m_{X_{D^*}}=2306$MeV and $\Gamma_{X_{D^*}}=21$MeV.  

Here it may be worthwhile to note that the 
preliminary values above obtained on 
preperties of $X_B$ and $X_{D^*}$ are consistent with 
our predictions by heavy quark symmetry and chiral symmetry, 
Eqs. (5) and (6).\ \ 
$ 
m_{X_B}-m_B = 261{\rm MeV} \approx m_{X_{D^*}}-m_{D^*}=296{\rm MeV},\ \ \ 
\Gamma_{X_B}=21{\rm MeV} = \Gamma_{X_{D^*}}=21{\rm MeV}.
$

\begin{figure}[t]
  \epsfxsize=14 cm
  \epsfysize=4.5 cm
 \centerline{\epsffile{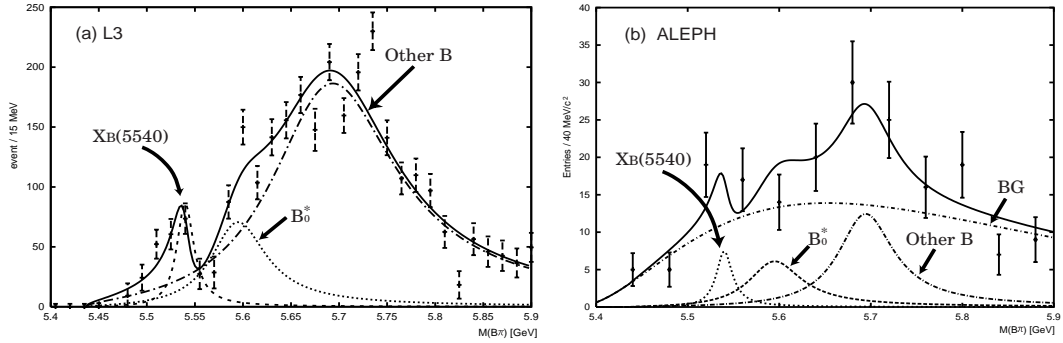}}
 \caption{$B\pi$ mass spectra obtained in (a) L3\cite{L3} and 
(b) ALEPH\cite{ALEPH} collaborations. 
Contributions from the individual Breit-Wigner amplitudes are shown by dotted lines.
In (a) the background contribution is subtracted.
The $m_{X_B}$ and $\Gamma_{X_B}$ are determined only through CLEO data, 
since ALEPH data have much less statistics.}
  \label{fig:1}
\end{figure}

\section{Concluding Remarks}
In the new classification-scheme,  
the chiral scalars $X_B,X_D$ and axial-vectors $X_{B^*},X_{D^*}$,
which are the chiral partners of $B,D$ and $B^*,D^*$, respectively, are
predicted to exist, which are to be discriminated 
from the conventional $P$-wave state mesons.
Preliminary analyses of 
experimental data give some indications of possible existence 
of both $X_B$ and $X_{D^*}$,
besides $B_0^*$ and $D_1^*$, with 
the properties consistent with the theoretical prediction.

This fact seems to suggest that the new level-classification scheme 
is actually realized in heavy-light quark meson systems.



\end{document}